# Dynamical prediction of two meteorological factors using the deep neural network and the long short-term memory (Ⅰ)

Ki-Hong Shin[a], Jae-Won Jung[c], Sung-Kyu Seo[c], Cheol-Hwan You[b], Dong-In Lee[b], Jisun Lee[b], Ki-Ho Chang[d], Woon-Seon Jung[d], Kyungsik Kim[a,c,*]

[a]*Department of Physics and* [b]*Department of Environmental Atmospheric Sciences, Pukyong National University, Busan 608-737, Republic of Korea*
[c]*DigiQuay Company Ltd., Seocho-gu Seoul 06552, Republic of Korea*
[d]*National Institute of Meteorological Research, Korea Meteorological Administration, Seogwipo 697-845, Republic of Korea*

**ABSTRACT**

It is important to calculate and analyze temperature and humidity prediction accuracies among quantitative meteorological forecasting. This study manipulates the extant neural network methods to foster the predictive accuracy. To achieve such tasks, we analyze and explore the predictive accuracy and performance in the neural networks using two combined meteorological factors (temperature and humidity). Simulated studies are performed by applying the artificial neural network (ANN), deep neural network (DNN), extreme learning machine (ELM), long short-term memory (LSTM), and long short-term memory with peephole connections (LSTM-PC) machine learning methods, and the accurate prediction value are compared to that obtained from each other methods. Data are extracted from low frequency time-series of ten metropolitan cities of South Korea from March 2014 to February 2020 to validate our observations. To test the robustness of methods, the error of LSTM is found to outperform that of the other four methods in predictive accuracy. Particularly, as testing results, the temperature prediction of LSTM in summer in Tongyeong has a root mean squared error (RMSE) value of 0.866 lower than that of other neural network methods, while the mean absolute percentage error (MAPE) value of LSTM for humidity prediction is 5.525 in summer in Mokpo, significantly better than other metropolitan cities.

---

* Corresponding authors. Fax: +82 51 629 4547.
E-mail addresses: kskim@pknu.ac.kr (K. Kim), leedi@pknu.ac.kr (D.-I. Lee).

# 1. Introduction

Through past decades, meteorologists and environmentalists have warned that the temperate phenomena change among atmospheric and climatic changes are coming to earth's crisis. Meteorological factors have resulted in considerable temporal-spatial variations for complex systems [1]. The statistical quantities of precipitation, heat transfer, solar radiation [2], temperature, wind, humidity, surface hydrology, and land subsidence [3] have been calculated and analyzed within better resolutions of each grid cell in our earth. With the weather prediction of the world meteorological organization (WMO), these correlations are presently proceeding to shed light on the atmospheric properties. In the past time, El Niño–southern oscillation forecast models have been categorized into three types: coupled physical models, statistical models, and hybrid models [4]. Among these models, the statistical models introduced for the ENSO forecasts have only been applied to the neural network model, multiple regression model, and canonical correlation analysis [5]. The meteorological statistical models have particularly made reasonable prediction accuracies in sea surface temperature anomalies [6]. Several researches have been simulated and analyzed the prediction accuracies of temperature and humidity, flood forecasting, flood warning, and natural hazard among quantitative meteorological forecasting. However, it is presently an important task for scientific researchers to extend and develop the forecast models using advanced neural network models.

Recently, the machine learning has been considerable attention in the field of physics such as statistical physics, particle physics, condensed matter physics, cosmology, and so on [7]. In particular, the investigation of statistical quantities on the Ising, XY, and Heisenberg models have been simulated and analyzed the prediction values in the restricted Boltzmann machine [8], restricted brief network, and recurrent models in the machine learning [9]. Barra et al. [10] have investigated a hybrid restricted Boltzmann machine with binary hidden and visible units. They showed the thermodynamics of visible units equivalent to a Hopfield network, and applied the method of stochastic stability to derive the thermodynamics in their model.

The artificial intelligence has been applied to various fields of pure and applied sciences, and such researches is actively making progress be ascertained the future prediction. Over past five decades, many predictive models have been proposed for human memory as a collective property of neural networks. The neural network models [11,12] have been based on a Hamiltonian extended by the equilibrium statistical mechanics, and the equilibrium properties of Hopfield model have in detail been discussed [13,14]. As well known, Werbos has furthermore proposed the backpropagation method for transitional artificial neural network (ANN) [15], and this is a method training the neural network by calculating the error between the output value in the forward direction and the actual value propagated the error in the reverse direction [16]. Researchers have introduced the gradient descent method to update weights by performing the neural network learning in the direction of minimizing errors [17].

Many studies have tried to minimize the error in the neural network models such as speech recognition, generation, handwriting and image recognitions, protein structure, and neurobiological systems from the viewpoint on the neural network method [18-21], and the learning effect from the viewpoint on feedforward neural networks has been obtained with a statistical–mechanical framework. The neural network algorithm has been concentrated on optimizing a suitable cost function on a class of network weights that quantifies and qualifies on the training and testing sets [22]. Indeed, meteorological researchers have successfully applied forecasting to the task of identifying patterns in meteorological time- series. On the other hand, the deep neural network (DNN) has its capability to extract useful features from high frequency time-series data, which can be used for estimation and prediction. The DNN algorithm in signal and image processing have been suggested a novel potential for enhancing the accuracy of precipitation estimation, when merging information from multiple earth-orbit satellite channels. Tao et al. [23] have studied a DNN for precipitation estimation using the satellite information, infrared, and water vapor channels, and they have particularly showed a two-stage framework for precipitation estimation from the meteorological information.

As estimating and analyzing the price, exchange rate, and bitcoin time-series for technical trading rules and methods in economics and econometrics, researchers have applied neural network architectures such as the ANN [24], recurrent neural network, psi-sigma neural network, and hybrid neural network models to the task of forecasting [25-26]. Sermpinis et al. [27] have recently applied traditional statistical prediction techniques to the ANN, recurrent neural network, and psi-sigma neural network for the EUR/USD exchange rate. They analyzed that the RMSE has a lower value in the case of neural network models. Tsaih et al. [28] have attempted to forecast stock index futures, and they showed that the cost function of neural network outperforms that of artificial neural network. Hussain et al. [29] have particularly analyzed the forecasting result of psi-sigma neural network for some exchange rates using univariate series as inputs. Dunis et al. [30] have studied exchange rate series with the psi-sigma neural network, but the psi-sigma neural network did not outperform better than the high order neural networks in a trading application. In mathematical and financial models, the ANN has furthermore been extended to the scientific hybrid models such as the neural network-genetic and neural network-fuzzy models. Integrated neural network models (the neural network-GARCH, neural network-EGARCH, and neural network-EWMA models) have particularly enhanced a predictive power by comparing the volatilities of single models, and these models have proposed and simulated in the framework of accuracy [31-33]. The prediction accuracy was calculated the small, medium, and large scale data by applying deep learning with the autoencoder, restricted Boltzmann machine, extreme learning machine (ELM), and radial basis function neural network [34]. Supervised learning includes regressions such as the linear regression, logistic regression, ridge regression, and Lasso regression, and the support vector machine and the decision tree [35-37]. On the other hand, the reinforcement learning, which is known as a learning with actions and rewards, exists to supervised and unsupervised learning [38,39].

Furthermore, the DNN has been widely and variously applied and treated in many scientific fields, such as signal and image processing, computer vision, language, and feature extraction [41]. In recent researches, the DNN techniques have proven to be influent and powerful when dealing with classification, prediction, and application problems [42, 43]. In addition to the optimizers for the optimization such as the Nestrov, AdaGrad, RMSProp, Adam, and AdamW [44-46], the DNN models have been applied to some fields such as the financial markets [47-50], transportations [51-56], weathers [57-61], voice recognitions [62-65], and electrical problems [66-68].

Indeed, Huang et al. [69] have proposed a new learning algorithm called the ELM for single-hidden layer feedforward neural network. The ELM [70] is shown to provide a good generalized performance learning thousands of times faster than conventional neural network model, and their results accomplished a good performance based on benchmark function approximation and classification problems [71-73]. Huang and Chen [74] have proved the universal approximation capability from the theoretical point of view via the ELM method, differently from many other learning methods. Savojardo et al. [75] have improved several computational methods to discriminate Transmembrane β-barrels from other types of proteins, and they showed the best performing approaches with a high fraction of false positive predictions.

The long short-term memory (LSTM) [76] has demonstrated a more superior method than transitional recurrent neural network in presence of long time lags. Gers and Schmidbuber [77] have introduced a novel long short-term memory with peephole connections (LSTM-PC) to overcome a weakness in the LSTM architecture. They showed that the LSTM translates the change of target signal into pertinent sequence-generating algorithm incorporated in recurrent connections [78-81]. Chen et al. have designed, trained and tested a convolutional LSTM encoder–decoder, and their approach was compared with LSTM-based methods using five crowded video sequences with public datasets. They also showed that their method reduces the displacement offset error and provides the realistic trajectory prediction [82].

Even more importantly, Moustra et al. have introduced a ANN model to predict the intensity of earthquakes in Greece in seismic neural network model. They have used a multilayer perceptron for both seismic intensity time-series data and seismic electric signals as input data [83]. In meteorological works, Tao et al. [84,85] have provided promising performances of the application of DNNs on precipitation among meteorological factors. They have discussed the benefits of DNNs incorporating the Kullback–Leibler divergence for precipitation estimation and prediction. Gonzalez et al. [86] have used the recurrent neural network and LSTM models to predict the earthquake intensity in Italy with hourly-data. Kashiwa et al. predicted rain-autumn for the local regions in Japan. They were applied the hybrid algorithm in the random optimization method [87]. The Convolution Neural Network (CNN) has used for high-level problems including the image recognition, object detection, and language processing [88,89,90]. Zhang and Dong have studied the CNN to predict the temperature by using the daily temperature data of China for learning data of seventy years [91].

To the best of our knowledge, there is of crucial value for investigating and analyzing the statistical forecasting from the understanding of relationship between temperature and humidity among meteorological predictors. There are not until now more current researches for meteorological factors with high and low frequency data. We only limit to the analysis of low frequency time-series data in this paper, and high frequency data will be turned to the next open study. The purpose of this paper is to study and analyze the dynamical prediction of two metrological factors using the neural network methods. We train and predict accuracies of average temperature and humidity for ten metropolitan cities in South Korea by applying five neural network models. Simulated studies are performed by applying the ANN, DNN, ELM, LSTM, and LSTM-PC machine learning methods and accuracy values are compared for each other ones. Data are extracted from low frequency time-series of ten metropolitan cities of South Korea, during seven years from March 2014 to February 2020 to validate our observations. Through the computer-simulation, the five neural network models are analyzed the testing results for 2500, 5000, and 7500 epochs and obtained the predicted accuracy of average temperature and humidity. In Section 2, the extracted data for calculations are given. The ANN, DNN, ELM, LSTM, and LSTM-PC are briefly characterized, and the recalled algorithms for their computation are outlined. Corresponding calculation and its result are presented for the data sets and parameters of the employed models in Section 3. A brief summary and major conclusion are gathered in Section 4.

# 2. Theoretical background

## 2-1. Data

Daily (low frequency) time-series data set of average temperature and humidity used in this study are taken from the Korea Meteorological Administration (KMA) database. The ten metropolitan cities we calculated and analyzed are Seoul, Incheon, Daejeon, Daegu, Busan, Pohang, Tongyeong, Gwangju, Mokpo, and Jeonju. We extract the time-series data of the manned regional meteorological offices of the KMA to ensure the reliability of data. We use the low frequency time-series data for seven years from 2014 to 2020. In this study, it consists of ~85% (of entire) data, from March 2014 to February 2019, to train the neural network models and remained ~15% (of entire) data, from March 2019 to February 2020, to test the predictive accuracy of the methods for two meteorological factors (temperature and humidity). We also use and calculate daily time-series data for the four seasons divided into the spring (March, April, May), the summer (June, July, August), the autumn (September, October, November), and the winter (December, January, February). Through the computer-simulation, the five neural network models are also analyzed the testing results after training for 2500, 5000, and 7500 epochs and obtained the predicted accuracy of temperature and humidity.

## 2-2. Prediction models

In this subsection, we introduce the method and its technique for neural network models, that is, the artificial neural network (ANN), deep neural network (DNN), extreme learning machine (ELM), long short-term memory (LSTM), and long short-term memory with peephole connections (LSTM-PC).

ANN is a mathematical machine learning method that presents some features of brain functions as a computer-simulation. That is, it is an artificially explored network, distinguished from a biological neural network. The neural network architectures are established to compare the ANN performance. When constructing the ANN model, data variables are normalized to the interval between 0 and 1 as follows: $x_{\text{norm}} = (x - x_{\text{min}})/(x_{\text{max}} - x_{\text{min}})$. The ANN structure has one input layer with four nodes and one hidden layer with three nodes. Our ANN structure is as follows: In four input nodes of one input layer, $T_{t-1}$ ($T_t$) denotes the temperature value at time lag $t-1(t)$, while $H_{t-1}$ ($H_t$) the humidity one at time lag $t-1(t)$. $HL_1$, $HL_2$, and $HL_3$ are three hidden nodes in the middle (hidden) layer. $T_{t+1}$ ($H_{t+1}$) denotes the output node of temperature (humidity) prediction at time lag $t$ +1. Hence, the value $T_{t+1}$ ($H_{t+1}$) is able to be found the predictive accuracy of temperature (humidity) in the output layer after learning iteratively [92].

The ANN with two or more hidden layers is called a DNN. Furthermore, in our study, wed construct a deep neural network (DNN) as one input with four nodes and two hidden layers with each three nodes, as shown in Fig. 1. That is, In input layer, $T_{t-1}$ , $T_{t,}$ , $H_{t-1,}$ and $H_t$ means the four input nodes for temperature and humidity at time lag $t$−1 and $t$, respectively. In first (second) hidden, layer, $HL_1^1$, $HL_2^1$, and $HL_3^1$ ($HL_1^2$, $HL_2^2$, and $HL_3^2$) are the hidden nodes. $T_{t+1}$ ($H_{t+1}$) denotes the output node of temperature (humidity) prediction at time lag $t$ +1.

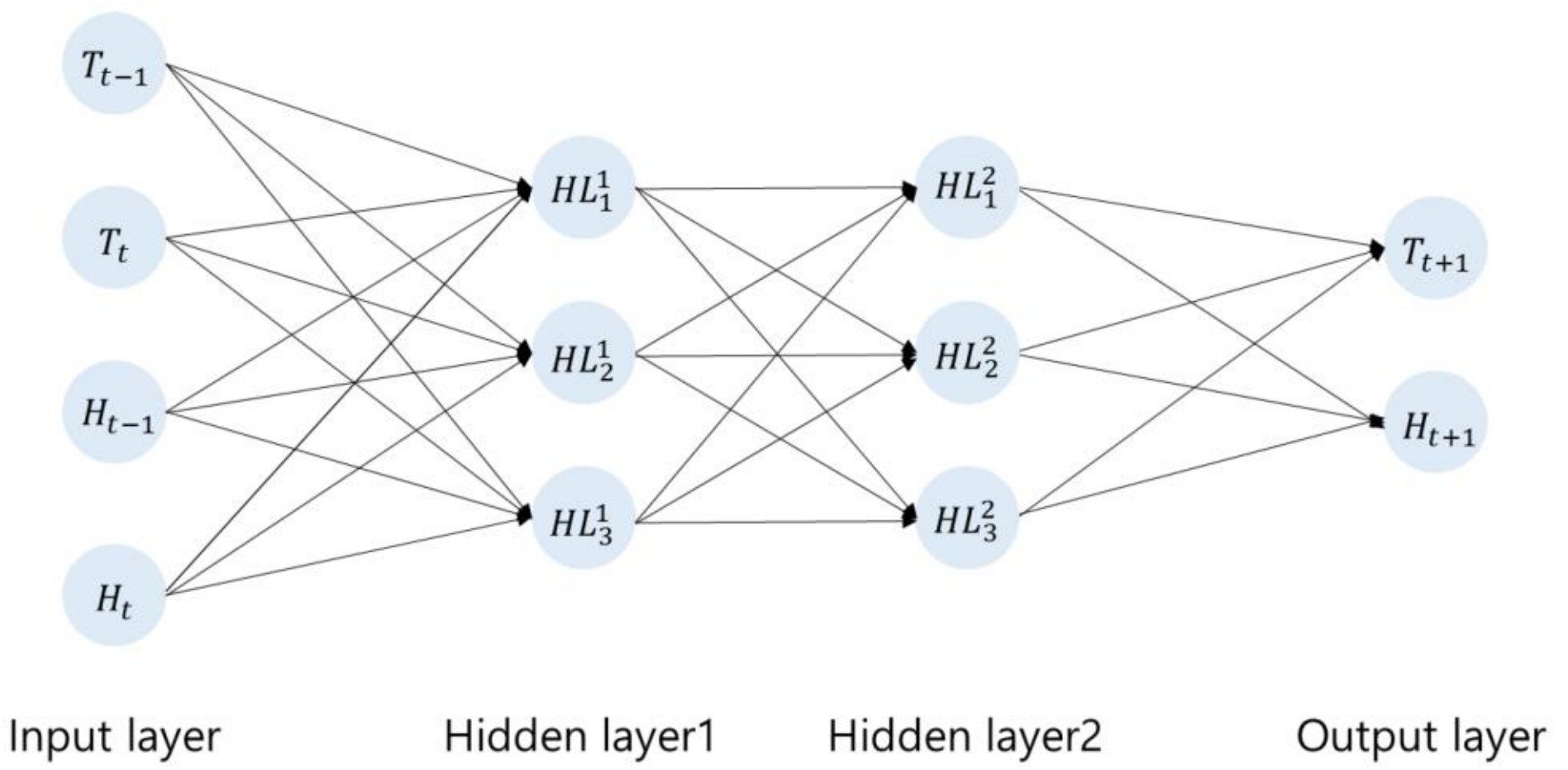

**Fig. 1:** Structure of two-hidden layered deep neural network (DNN). Here, the superscript indicates the number of hidden layers, and the subscript indicates the number of nodes.

From the calculation of ANN and DNN, the weighted sum of an external stimulus entered into the input layer is output as the pertinent reaction through the activation function. That is, the weighted sum of the external stimulus is represented in terms of

$$y_j = \sum_{i=1}^{n} w_{ij}\, x_i, \tag{1}$$

where $y_j$ denotes the weighted sum of external stimulus entered into the $j$-th output,

$x_i$ the external stimulus, and $w_{ij}$ the connection weight of output . As the reaction value of output is determined by the activation function, the sigmoid function for the analog output is used as follows:

$$\sigma(y_j) = 1/[1 + \exp(-\beta y_j)] \ . \tag{2}$$

Here, we generally use $\beta$ = 1 in order to avoid the divergence of σ($y_j$) and to obtain the optimal value of the connection weight of output. The learning of the counter propagation algorithm can minimize the sum of errors, as compared with the calculated value of all directional feed forward to the target value. After we continuously adjust the connection weight, we can iterate the learning process. Next, in order to evaluate predictive accuracy, we can use four criteria as follows: the root mean square error (RMSE), mean absolute percentage error (MAPE), mean absolute error (MAE), and Theil's − U [27] are presented in terms of

$$\text{RMSE} = \sqrt{\frac{1}{N}\sum_{t=1}^{N}(y_t - \hat{y}_t)^2} \tag{3}$$

$$\text{MAPE} = \frac{1}{N}\sum_{t=1}^{N}\left|\frac{y_t - \hat{y}_t}{y_t}\right| \tag{4}$$

$$\text{MAE} = \frac{1}{N}\sum_{t=1}^{N}|y_t - \hat{y}_t| \tag{5}$$

$$\text{Theil's U} = \frac{\sqrt{\frac{1}{N}\sum_{t=1}^{N}(y_t - \hat{y}_t)^2}}{\sqrt{\frac{1}{N}\sum_{t=1}^{N} y_i^2} + \sqrt{\frac{1}{N}\sum_{t=1}^{N} \hat{y}_i^2}} \ . \tag{6}$$

where $y_t$ and $\hat{y}_t$ are actual value and predicted value at time lag $t$, respectively, and $N$ is the data size of the tested set. The RMSE represents the standard deviation of the difference between the actual and predicted values in each neural network model. The MAPE also means the accuracy as a percentage, and the MAPE denotes the mean absolute relative error in each neural network model. As well-known, the smaller the values of RMSE and MAPE, the higher the accuracy of the model.

The extreme learning machine (ELM) is a feedforward neural network for classification, regression, and clustering. Let us recall the ELM method [69,70], and this is a feature learning with a single layer of hidden nodes. Using a set of training samples $\left[(x_j, y_j)\right]_{j=1}^{s}$ for $s$ samples and $v$ classes, the activation function $g_i(x_j)$ for the single hidden layer with $n$ nodes is given by

$$z_j = \sum_{i=1}^{n} \beta_i\, g_i(x_j) = \sum_{i=1}^{n} \beta_i\, g_i(w_i \cdot x_j + b_i),\ j = 1,2,\dots,s. \quad (7)$$

Here, $x_j = [x_{j1}, x_{j2}, \dots, x_{js}]^{\mathrm{T}}$ is the input units, and $y_j = [y_{j1}, y_{j2}, \dots, y_{jv}]^{\mathrm{T}}$ the output units. The statistical quantity $w_i = [w_{j1}, w_{j2}, \dots, w_{js}]^{\mathrm{T}}$ denotes the connecting weights of hidden unit *i* to input units, $b_i$ the bias of hidden unit $i$, $\beta_i = [\beta_{i1}, \beta_{i2}, \dots, \beta_{iv}]^{\mathrm{T}}$ the connecting weights of hidden unit *i* to the output units, and $z_j$ the actual network output. The ELM method can solve using error minimization as $\min_\beta ||H\beta - z||_f$ with

$$H(\boldsymbol{w}_1, \dots, \boldsymbol{w}_n, b_1, \dots, b_n) = \begin{bmatrix} g(\boldsymbol{w}_1 \cdot \boldsymbol{x}_1 + b_1) & \cdots & g(\boldsymbol{w}_{\tilde{N}} \cdot \boldsymbol{x}_1 + b_n) \\ \vdots & \cdots & \vdots \\ g(\boldsymbol{w}_1 \cdot \boldsymbol{x}_s + b_1) & \cdots & g(\boldsymbol{w}_{\tilde{N}} \cdot \boldsymbol{x}_s + b_n) \end{bmatrix}_{n \times s} \quad (8)$$

and

$$z = [z_1^T, z_2^T, \dots, z_s^T]^{\mathrm{T}}. \quad (9)$$

Here, $H$ is the output matrix of hidden layer, and $\beta$ the output weight matrix. The ELM randomly selects the hidden unit parameters, and the output weight parameters need to be determined.

For the LSTM model, Hochreiter and Schmidhuber [76] have proposed the LSTM technique in order to solve the long-term dependence problem of common RNNs. Particularly, the recurrent neural network is known as a generalized and specialized algorithm for processing long-time series data, and this is possible to learn by using the novel input data from subsequent step and the output data from the previous step at the same time. The LSTM is known to be advantageous both for preventing the inherent vanishing gradient problem of general recurrent neural network and for predicting time series data [76,77].

Let us recall the LSTM is composed of a cell with three gates attached as follows. As well-known, these are divided into forget, input, and output gates to protect and control the cell state. Let us recall one cell of LSTM. As the first step in the cell, the forget gate $f_t$ (input gate $i_t$) enters the input value $x_t$ and an output value $h_{t-1}$ through the previous step, where $x_t$ and $h_{t-1}$ are the normalized values between zero and one. If the output information is exited, $f_t$ is represented in terms of

$$i_t = \sigma(w_{xi}x_t + w_{hi}h_{t-1} + b_i) \tag{10}$$

$$f_t = \sigma(w_{xf}x_t + w_{hf}h_{t-1} + b_f) \ . \tag{11}$$

Here, $\sigma(y)$ denotes the activation function as a function of y, $w_{xi}$ and $w_{xi}$ the weights of gate, and $b_i$ and $b_f$ an bias value. In the second step, the input gate has the new cell states updated, when subsequent values are entered as follows:

$$\overline{c_t} = \tanh( w_{xc}x_t + w_{hc}h_{t-1} + b_c) \tag{12}$$

$$c_t = f_t \odot c_{t-1} + i_t \odot \overline{c_t} \ . \tag{13}$$

Here, $c_t$ is the cell state, and $\overline{c_t}$ the activation function created through the output gate. $w_{xc}$ and $w_{hc}$ are the weight values in output gate, and $b_c$ the bias value. The symbol $\odot$ represents the inner product of matrices. In the third step, the output gate $o_t$ is exited as

$$o_t = \sigma(w_{xo}x_t + w_{ho}h_{t-1} + b_o) \ . \tag{14}$$

Here, this gate is determined an output value. Lastly, an new predicted output value $h_t$ is calculated as

$$h_t = o_t \odot \tanh c_t \ . \tag{15}$$

The LSTM has previously introduced as a novel type of recurrent neural network, and this method performs better than the recurrent neural network for long time lags. The architecture of LSTM have played a crucial role in connecting long time lags between input events.

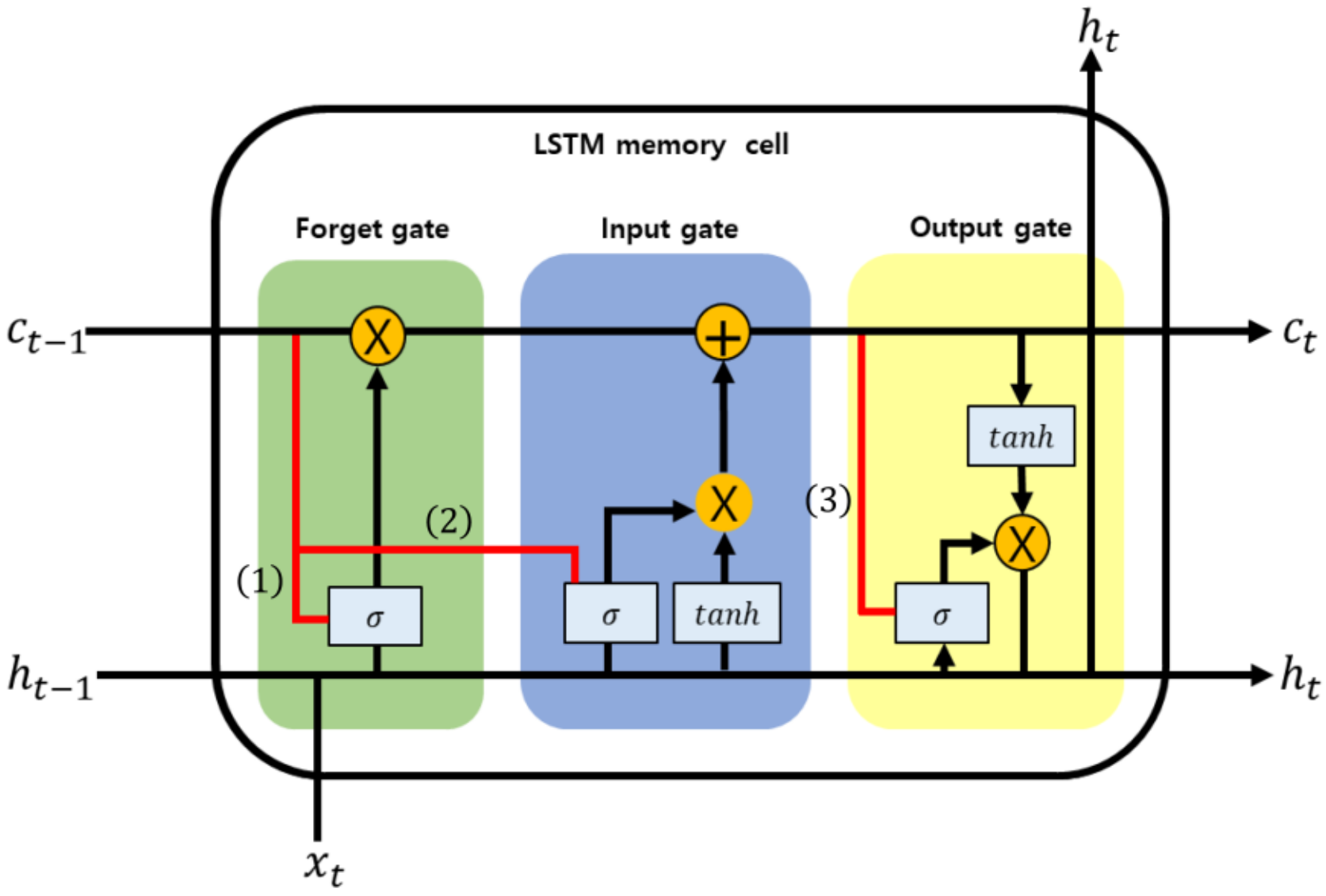

**Fig. 2:** LSTM-PC structure including the paths (1)- (3). Here, the path (1) ((2)) is one peephole connection connected from $c_{t-1}$ to forget gate (input gate), and the path (3) is one peephole connection connected from $c_t$ to output gate.

The RNN has in principle learned to make use of numerous sequential tasks such as motor control and rhythm detection. It is well known that the LSTM outperforms other RNNs on tasks involving long time lags. Let us recall the LSTM with peephole connections (LSTM-PC) method added the three paths (1)-(3) rather than the LSTM, as shown in Fig. 2. This method augmented from the internal cells to the multiplicative gates can take the fine distinction between sequences of spikes [74]. The LSTM-PC cell is also made up as the forget gate, input gate, and output gate.

The peephole connections allows the gates to carry out their operations as a function of both the incoming inputs and the previous state of the cell. In Fig. 2, the LSTM-PC implements the compound recursive function to obtain the predicted output value $h_t$ as follows:

$$i_t = \sigma(w_{xi}x_t + w_{hi}h_{t-1} + w_{ci}c_{t-1} + b_i) \tag{16}$$
$$f_t = \sigma(w_{xf}x_t + w_{hf}h_{t-1} + w_{cf}c_{t-1} + b_f) \tag{17}$$
$$c_t = f_t \odot c_{t-1} + i_t \odot \tanh(w_{xc}x_t + w_{hc}h_{t-1} + b_c) \tag{18}$$
$$o_t = \sigma(w_{xo}x_t + w_{ho}h_{t-1} + w_{co}c_t + b_o) \tag{19}$$
$$h_t = o_t \odot \tanh c_t \ , \tag{20}$$

where $i_t$, $f_t$, $c_t$, and $o_t$ are the input gate, forget gate, cell gate, and output gate activation vectors at time lag $t$, respectively, and $w_{ci}$, $w_{cf}$, and $w_{co}$ the peephole weights. In view of Eqs. (16), (17), and (19), we can discriminate the LSTM-PC from the LSTM. From Fig. (2), the path (1) ((2)) has the value of $w_{ci}c_{t-1}$ $(w_{cf}c_{t-1})$ added in one peephole connection connected from $c_{t-1}$ to forget gate (input gate). In path (3), $w_{co}c_t$is the value added in one peephole connection connected from $c_t$ to output gate. It will be particularly showed in section 3 that the LSTM and LSTM-PC are set for different three training sizes over 2500, 5000, and 7500 epochs.

There is clearly a need for searching a peculiar neural network method that is more superior than the other neural networks for the predictive accuracy of meteorological factors. It has been evaluated that the ANN method has been extended and developed into hybrid models by combining with different genes, fuzzy, and wavelet models, and has positively improved the accuracy of prediction. Among some methods [92], it is considered that the DNN should be developed by increasing and controlling more suitably the number of hidden layers. We anticipate that the LSTM and the LSTM-PC have excellent results of reducing error if the learning rate value and the number of epochs are adequately tried and regulated for an specific testing problem.

# 3. Numerical calculations and results

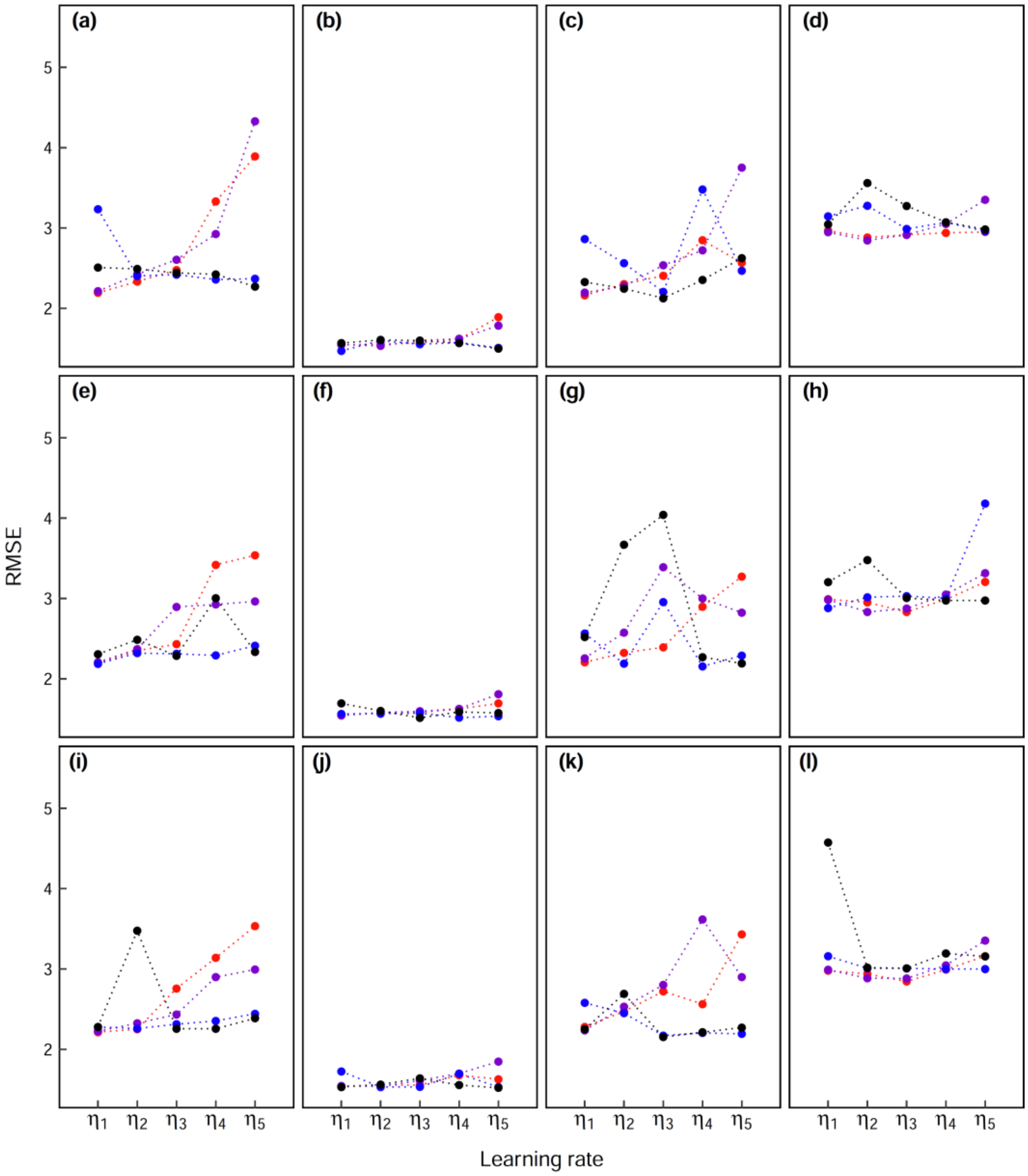

**Fig. 3:** Values of the RMSE as a function of the learning rate in the ANN (red circle), DNN (purple circle), LSTM (blue circle), and LSTM-PC (black circle) in all seasons of Seoul in testing 1. Here, (a)-(d) , (e)-(h), and (i)-(l) are, respectively, the results for training 2500, 5000, and 7500 epochs in the spring, summer, autumn, and winter.

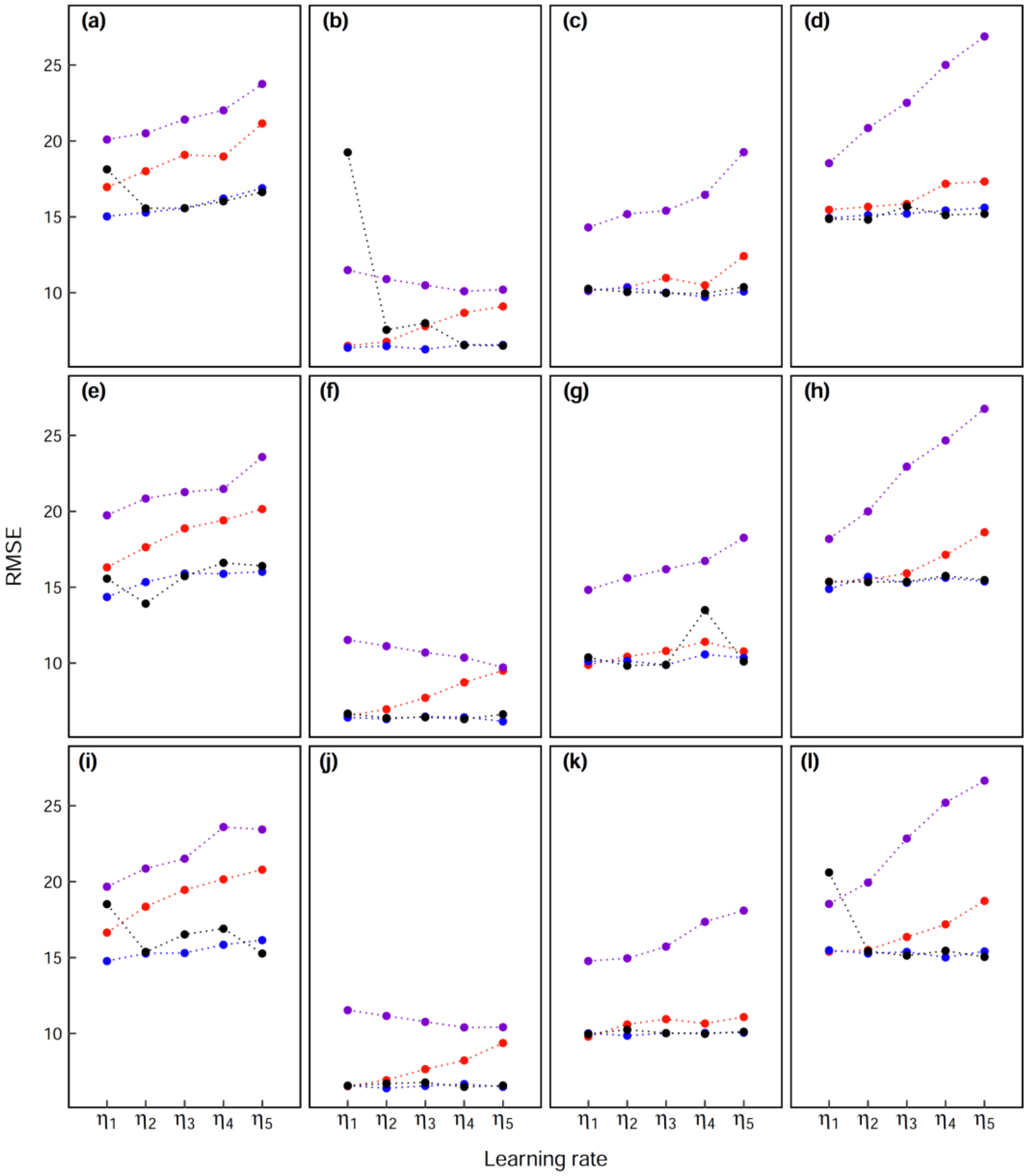

**Fig. 4:** Values of the MAPE as a function of the learning rate in the ANN (red circle), DNN (purple circle), LSTM (blue circle), and LSTM-PC (black circle) in all seasons of Tongyeong in testing 2. Here, (a)-(d), (e)-(h), and (i)-(l) are, respectively, the results for training 2500, 5000, and 7500 epochs in the spring, summer, autumn, and winter.

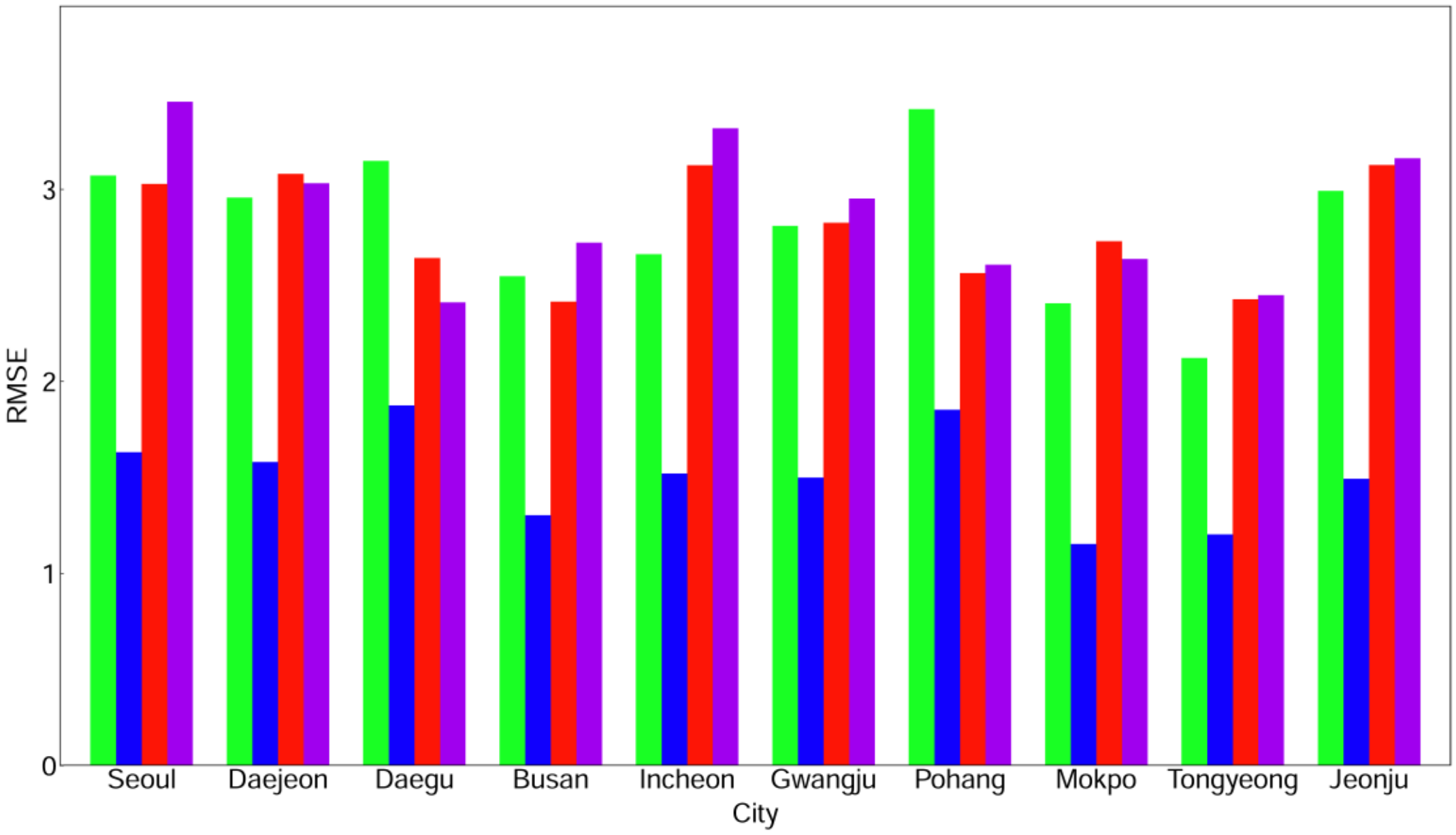

**Fig. 5:** The RMSE of ELM for 7500 epochs in spring (green bar), summer (blue bar), autumn (red bar), winter (purple bar) in testing 1.

**Table 1.**
RMSE values of ELM for 7500 epochs for spring, summer, autumn, and winter in ten metropolitan cities in testing 1, where the numbers in parentheses indicate the learning rate $\eta$.

| | Seoul | Daejeon | Daegu | Busan | Incheon | Gwangju | Pohang | Mokpo | Tongyeong | Jeonju |
|---|---|---|---|---|---|---|---|---|---|---|
| **Spring** | 2.259 (0.003) | 2.216 (0.003) | 2.54 (0.007) | 1.973 (0.003) | 1.935 (0.003) | 2.16 (0.005) | 2.884 (0.005) | 1.903 (0.003) | 1.629 (0.005) | 2.355 (0.007) |
| **Summer** | 1.529 (0.009) | 1.356 (0.003) | 1.58 (0.001) | 1.003 (0.001) | 1.352 (0.009) | 1.281 (0.003) | 1.554 (0.005) | 0.935 (0.003) | 0.898 (0.003) | 1.283 (0.007) |
| **Autumn** | 2.171 (0.005) | 2.317 (0.007) | 1.852 (0.009) | 1.832 (0.003) | 2.259 (0.007) | 2.107 (0.009) | 1.824 (0.003) | 1.995 (0.005) | 1.786 (0.009) | 2.281 (0.005) |
| **Winter** | 2.999 (0.007) | 2.606 (0.007) | 2.124 (0.009) | 2.522 (0.009) | 2.801 (0.009) | 2.485 (0.009) | 2.335 (0.009) | 2.368 (0.005) | 2.248 (0.005) | 2.686 (0.007) |

In this section, the computer-simulation is performed for testing 1 and 2 as follows: testing 1 has the four nodes $T_{t-1}$, $T_t$, $H_{t-1}$, $H_t$, in the input layer and the one output node $T_{t+1}$in output layer, and testing 2 has also the four input nodes $T_{t-1}$, $T_t$, $H_{t-1}$, $H_t$, and the one output node $H_{t+1}$. It is not known beforehand what values of learning rates are appropriate. However, we select the five learning rates of $\eta$ =0.1, 0.2, 0.3, 0.4, and 0.5 for the ANN and the DNN, while the learning rate values for LSTM and LSTM-PC are $\eta$ =0.001, 0.003, 0.005, 0.007, and 0.009, for different training sizes over three runs, 2500, 5000, and 7500 epochs. Particularly, the predicted accuracies of ELM are also obtained by averaging the results over 2500, 5000, and 7500 epochs.

Fig. 3 shows the predicted values of the RMSE as a function of the learning rate $\eta$ in the ANN, DNN, LSTM, and LSTM-PC in all seasons of Seoul in testing 1. Here, Figs. 3(a)-3(d), 3(e)-3(h), and 3(i)-3(l) are, respectively, the results for training 2500, 5000, and 7500 epochs in the spring, summer, autumn, winter.

From Fig. 3(a) for 2500 epochs in spring in Seoul, the ANN and the DNN show a tendency to increase the RMSE as each learning rate increases from $\eta$=0.1 to 0.5. The RMSE value of ANN is 2.192 at $\eta$=0.1, and has a high value of 3.82 at $\eta$=0.5. The RMSE value of DNN is 2.212 (4.33) at $\eta$=0.1 (0.5). Then, the RMSE value gradually increases as each learning rate increases from $\eta$=0.1 to 0.5. In Fig. 3(e) for 5000 epochs in summer in Seoul, the ANN and the DNN show a tendency to increase the RMSE as the learning rate increases from $\eta = 0.1$ to 0.5. The ANN RMSE value is 2.193 at $\eta = 0.1$, and has a high value of 3.538 at $\eta = 0.5$. The DNN RMSE value is 2.207 (2.963) at $\eta = 0.1$ (0.5). The RMSE of LSTM and LSTM-PC RMSEs do not show a clear trend. However, the ANN exhibits a higher RMSE value at $\eta$=0.5 compared to other learning rates. The RMSE of LSTM has significantly a value of 3.234 (2.187) for training 2500 (5000) epochs at $\eta = 0.1$. In Fig. 3(i) for 5000 epochs, the ANN and DNN shows a tendency to increase RMSE as the learning rate increases from $\eta = 0.1$ to 0.5. The RMSE value of ANN (DNN) is 2.214 (2.226) at $\eta$=0.1, and has a highest value of 3.535 (2.993) at $\eta_5$=0.5. The RMSE value (3.476) of LSTM-PC at $\eta = 0.2$ is larger than other learning rates. The RMSE of LSTM has a lower (higher) value of 2.259 (2.442) at $\eta = 0.2$ (0.5).

In Fig. 3(b), the RMSE of ANN and DNN has larger values than those of LSTM and LSTM-PC for 2500 epochs in summer in Seoul. The RMSE of ANN (DNN) value is 1.89 (1.785) at $\eta$=0.5, and the RMSE of LSTM (LSTM-PC) has a lower value of 1.507 (1.498) at $\eta$=0.5. In addition, as the learning rate increases, the RMSE value of ANN tends to increase. The RMSE of ANN had a lowest (highest) value of 1.545 (1.89) at $\eta$=0.1 (0.5). In Fig. 3(f) for training 5000 epochs, the RMSE of DNN has a highest value of 1.548 at $\eta$=0.1, but the RMSEs of ANN, LSTM, and LSTM-PC do not show a distinct trend. In Fig. 3(j) for 5000epochs, the RMSE of ANN shows very similar result with 1.554 (1.555) at $\eta$=0.2 (0.3). the LSTM-PC showed a higher RMSE value at $\eta$=0.3, compared to other learning rates.

Fig. 3(c) for 2500 epochs in autumn in Seoul, the RMSE of DNN tends to increase as the learning rate increases. The RMSE value of DNN has a lowest (highest) value of 2.195 (3.75) at $\eta$=0.1 (0.5). In the LSTM-PC, the RMSE decreased and then increased again based on $\eta$=0.3. In Fig. 3(g) for 5000 epochs, the RMSEs of DNN, LSTM, and LSTM-PC had highest values of 3.39, 2.956, and 4.04 at $\eta = 0.3$, respectively. The RMSE of ANN had a lowest (highest) value of 2.205 (3.273) at $\eta = 0.1$ (0.5). In Fig. 3(k) for 7500 epochs, the RMSE values of LSTM and LSTM-PC are 2.171 and 2.153 at $\eta = 0.3$, respectively. and it was confirmed that learning was better than the result of Fig. 3(g). The RMSE of ANN has a lowest (highest) value of 2.278 (3.431) at $\eta = 0.1$ (0.5). Hence, the learning rate gradually increases, the RMSE shows a tendency to increase.

Figs. 3(d), 3(h), and 3(l) are the results of learning 2500, 5000, and 7500 epochs for winter temperature. In Fig. 3(d), the RMSE of ANN and the DNN shows a tendency to increase from $\eta = 0.1$ to 0.5, while that of LSTM-PC shows a tendency to decrease from $\eta$=0.2 to $\eta_5$=0.5. The RMSE of LSTM do not show clearly a trend. In Fig. 3(h), the RMSE value of LSTM shows higher than that of other neural network methods at $\eta = 0.5$, and the RMSE value of LSTM-PC shows higher at $\eta = 0.2$ compared to other learning rates for each method. In Fig. 3(l), the RMSE of LSTM-PC shows a very higher value of 2.88, compared with those at $\eta = 0.1$ in Fig. 3(d)-3(h). On the other hand, at $\eta$=0.2, the RMSE of LSTM-PC shows better performance than those of Figs. 3(d) and 3(h).

In Fig. 4, we obtain the predicted values of the MAPE as a function of the learning rate in the ANN, DNN, LSTM, and LSTM-PC in all seasons of Tongyeong in testing 2. Here, Figs. 4(a)-4(d), 4(e)-4(h), and 4(i)-4(l) are, respectively, the results for training 2500, 5000, and 7500 epochs in the spring, summer, autumn, winter. Fig. 5 presents the RMSE of ELM for 7500 epochs in spring, summer, autumn, winter in testing 1, and Table 1 also illustrates the comparison of the RMSE of ELM for 7500 epochs for spring, summer, autumn, and winter in ten metropolitan cities in testing.

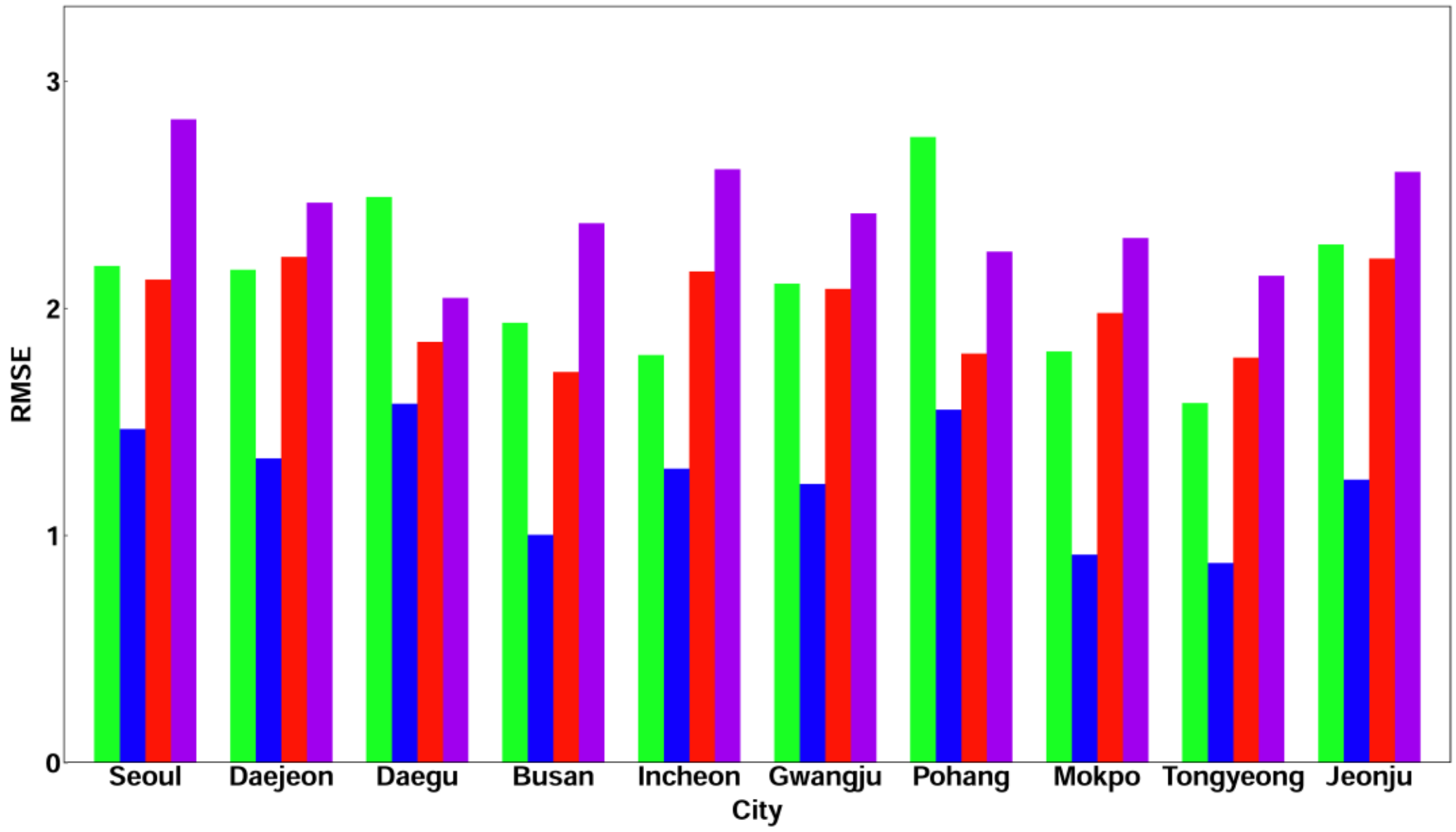

**Fig. 6:** The Lowest RMSE of ten metropolitan cities for all three training epochs in spring (green bar), summer (blue bar), autumn (red bar). winter (purple bar) in testing 1.

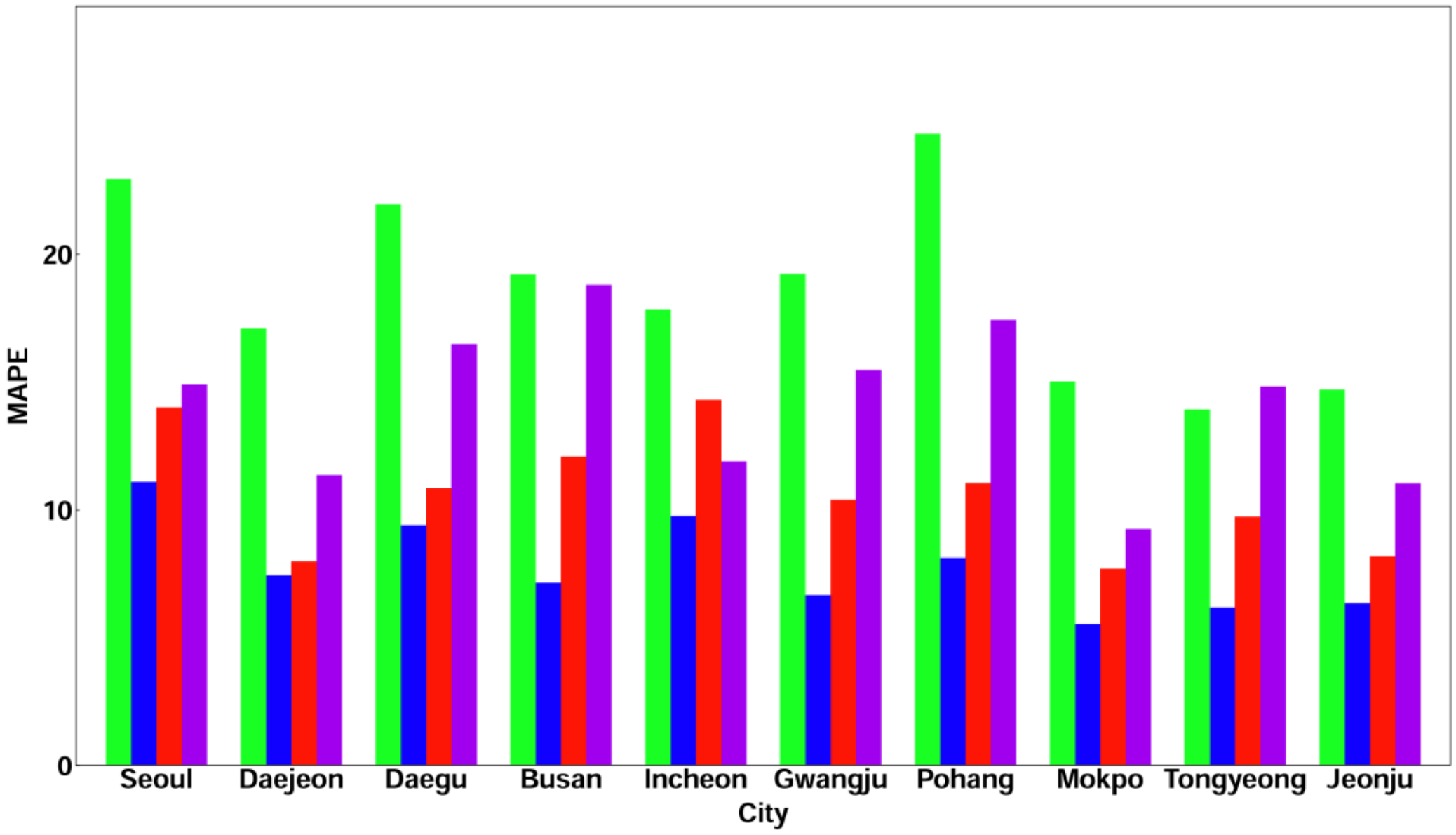

**Fig. 7:** The Lowest MAPE of ten metropolitan cities for all three training epochs in spring (green bar), summer (blue), autumn (red). winter (purple) in testing 2.

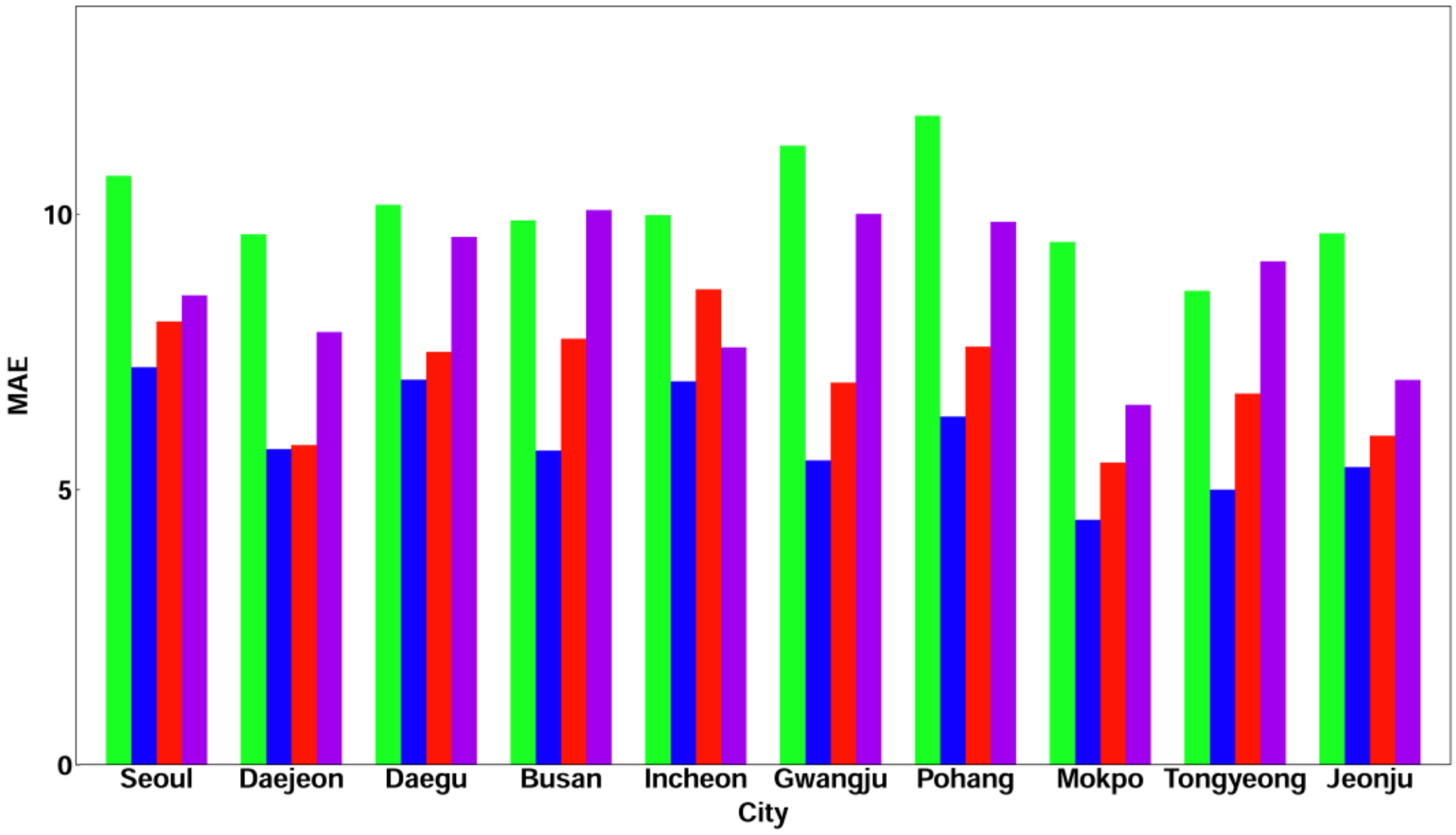

**Fig. 8:** The Lowest MAE of ten metropolitan cities for all three training epochs in spring (green bar), summer (blue bar), autumn (red bar). winter (purple bar) in testing 2.

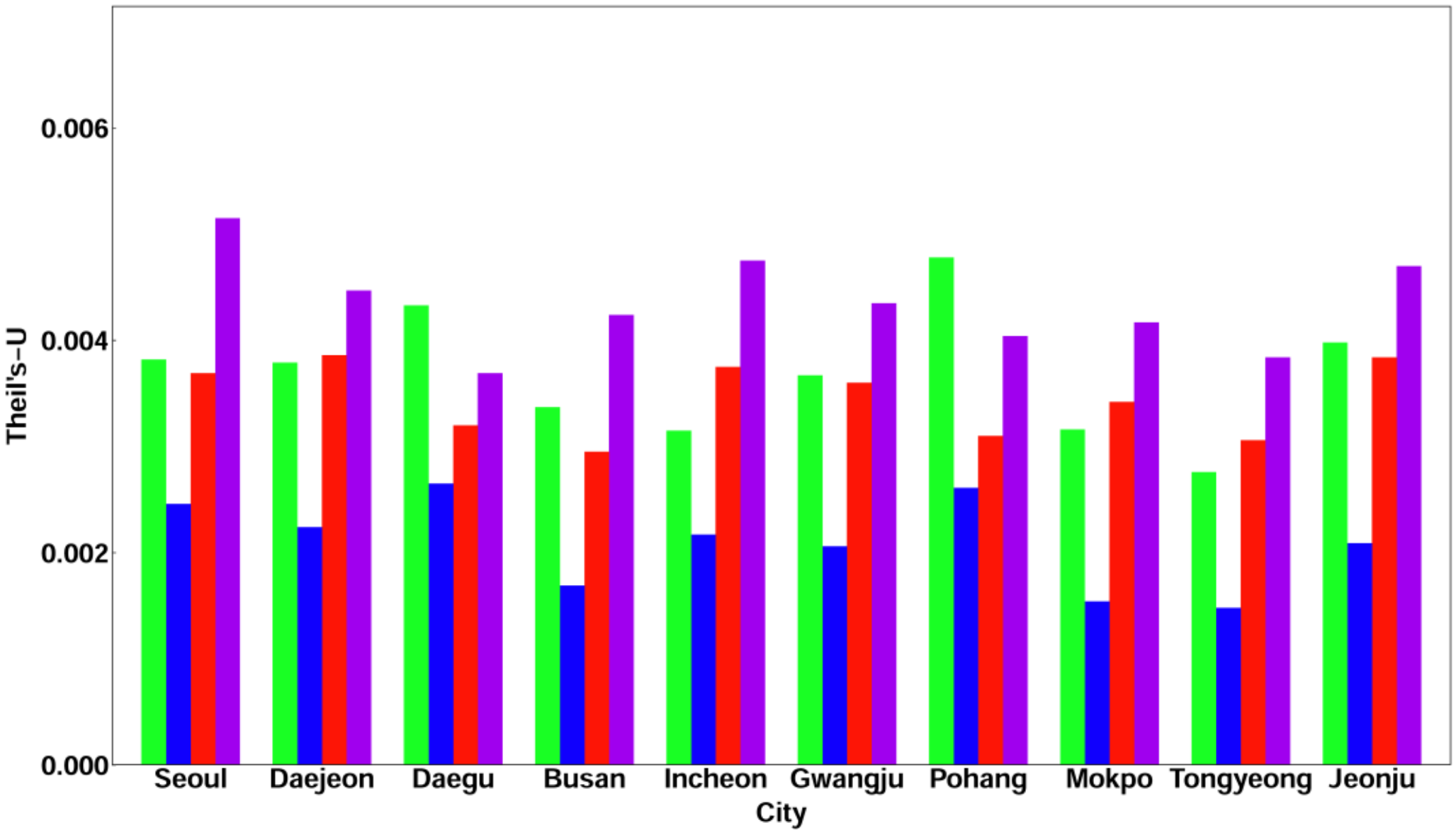

**Fig. 9:** The Lowest Theil's-U of ten metropolitan cities for all three training epochs in spring (green bar), summer (blue bar), autumn (red bar). winter (purple bar) in testing 1.

**Table 2.**
Values of RMSE, MAPE, MAE and Theil's-U of ten metropolitan cities in four seasons in testing 1, where $\eta$ denotes the learning rate.

| City | Season | RMSE | MAPE | MAE | Theil's-U ($\times 10^{-3}$) |
|---|---|---|---|---|---|
| Seoul | Spring | 2.187 (LSTM $\eta = 0.001$) | 0.61 (DNN $\eta = 0.1$) | 1.747 (DNN $\eta = 0.1$) | 3.823 (LSTM $\eta = 0.001$) |
| | Summer | 1.469 (LSTM $\eta = 0.001$) | 0.392 (LSTM $\eta = 0.001$) | 1.169 (LSTM $\eta = 0.001$), | 2.462 (LSTM $\eta = 0.001$) |
| | Autumn | 2.126 (LSTM-PC $\eta = 0.005$) | 0.392 (LSTM-PC $\eta = 0.009$) | 1.575 (LSTM-PC $\eta = 0.005$) | 3.686 (LSTM-PC $\eta = 0.005$) |
| | Winter | 2.832 (ANN $\eta = 0.5$) | 0.552 (LSTM-PC $\eta = 0.005$) | 2.046 (DNN $\eta = 0.3$) | 5.153 (ANN $\eta = 0.5$) |
| Daejeon | Spring | 2.17 (ANN $\eta = 0.1$) | 0.591 (DNN $\eta = 0.1$) | 1.694 (DNN $\eta = 0.1$) | 3.785 (ANN $\eta = 0.1$) |
| | Summer | 1.339 (LSTM $\eta = 0.001$) | 0.333 (LSTM $\eta = 0.007$) | 0.992 (LSTM $\eta = 0.007$) | 2.244 (LSTM $\eta = 0.001$) |
| | Autumn | 2.227 (LSTM-PC $\eta = 0.007$) | 0.58 (LSTM-PC $\eta = 0.007$) | 1.66 (LSTM-PC $\eta = 0.007$) | 3.855 (LSTM-PC $\eta = 0.007$) |
| | Winter | 2.465 (ANN $\eta = 0.5$) | 0.692 (ANN $\eta = 0.5$) | 1.909 (ANN $\eta = 0.5$) | 4.469 (ANN $\eta = 0.5$) |
| Daegu | Spring | 2.491 (LSTM-PC $\eta = 0.007$) | 0.677 (LSTM-PC $\eta = 0.009$) | 1.947 (LSTM-PC $\eta = 0.009$) | 4.328 (LSTM-PC $\eta = 0.007$) |
| | Summer | 1.58 (LSTM $\eta = 0.001$) | 0.407 (LSTM $\eta = 0.001$) | 1.209 (LSTM $\eta = 0.001$) | 2.646 (LSTM $\eta = 0.001$) |
| | Autumn | 1.852 (LSTM $\eta = 0.009$) | 0.491 (LSTM-PC $\eta = 0.007$) | 1.412 (LSTM-PC $\eta = 0.007$) | 3.202 (LSTM $\eta = 0.009$) |
| | Winter | 2.046 (ANN $\eta = 0.3$) | 0.545 (ANN $\eta = 0.3$) | 1.509 (ANN $\eta = 0.3$) | 3.694 (ANN $\eta = 0.3$) |
| Busan | Spring | 1.936 (DNN $\eta = 0.1$) | 0.522 (ANN $\eta = 0.3$) | 1.499 (ANN $\eta = 0.3$) | 3.366 (DNN $\eta = 0.1$) |
| | Summer | 1.003 (LSTM $\eta = 0.001$) | 0.266 (LSTM $\eta = 0.001$) | 0.789 (LSTM $\eta = 0.001$) | 1.686 (LSTM $\eta = 0.001$) |
| | Autumn | 1.72 (LSTM-PC $\eta = 0.005$) | 0.448 (LSTM-PC $\eta = 0.005$) | 1.298 (LSTM $\eta = 0.005$) | 2.955 (LSTM-PC $\eta = 0.005$) |
| | Winter | 2.375 | 0.64 | 1.793 | 4.241 |
| | | (DNN $\eta$ = 0.3) | (DNN $\eta$ = 0.1) | (DNN $\eta$ = 0.1) | (DNN $\eta$ = 0.3) |
| Incheon | Spring | 1.795 (ANN $\eta$ = 0.1) | 0.493 (DNN $\eta$ = 0.001)) | 1.405 (NN $\eta$ = 0.1)) | 3.149 (ANN $\eta$ = 0.1) |
| | Summer | 1.294 (LSTM $\eta$ = 0.009) | 0.334 (LSTM $\eta$ = 0.009) | 0.997 (LSTM $\eta$ = 0.009) | 2.173 (LSTM $\eta$ = 0.009) |
| | Autumn | 2.163 (LSTM $\eta$ = 0.00 | 0.569 (ANN $\eta$ = 0.3) | 1.628 (ANN $\eta$ = 0.3) | 3.746 (ANN $\eta$ = 0.3) |
| | Winter | 2.612 (LSTM-PC $\eta$ = 0.005) | 0.726 (DNN $\eta$ = 0.005) | 1.991 (DNN $\eta$ = 0.5) | 4.751 (LSTM-PC $\eta$ = 0.005) |
| Gwangju | Spring | 2.108 (LSTM-PC $\eta$ = 0.007) | 0.565 (ANN $\eta$ = 0.1) | 1.617 (ANN $\eta$ = 0.1) | 3.674 (LSTM-PC $\eta$ = 0.007) |
| | Summer | 1.227 (LSTM-PC $\eta$ = 0.001) | 0.314 (LSTM-PC $\eta$ = 0.001) | 0.934 (LSTM-PC $\eta$ = 0.1) | 2.058 (LSTM-PC $\eta$ = 0.001) |
| | Autumn | 2.086 (LSTM $\eta$ = 0.005) | 0.518 (LSTM-PC $\eta$ = 0.007) | 1.487 (LSTM-PC $\eta$ = 0.007) | 3.599 (LSTM $\eta$ = 0.005) |
| | Winter | 2.418 (LSTM-PC $\eta$ = 0.001) | 0.633 (ANN $\eta$ = 0.3) | 1.762 (ANN $\eta$ = 0.3) | 4.354 (LSTM-PC $\eta$ = 0.001) |
| Pohang | Spring | 2.754 (LSTM $\eta$ = 0.001) | 0.745 (LSTM $\eta$ = 0.009) | 2.147 (LSTM $\eta$ = 0.009) | 4.783 (LSTM $\eta$ = 0.001) |
| | Summer | 1.554 (LSTM $\eta$ = 0.005) | 0.436 (LSTM $\eta$ = 0.001 | 1.299 (LSTM $\eta$ = 0.001) | 2.606 (LSTM $\eta$ = 0.005) |
| | Autumn | 1.801 (ANN $\eta$ = 0.1) | 0.482 (LSTM-PC $\eta$ = 0.007) | 1.393 (LSTM-PC $\eta$ = 0.007) | 3.101 (ANN $\eta$ = 0.1) |
| | Winter | 2.249 (DNN $\eta$ = 0.1) | 0.598 (DNN $\eta$ = 0.1) | 1.666 (DNN $\eta$ = 0.1) | 4.041 (DNN $\eta$ = 0.1) |
| Mokpo | Spring | 1.81 (LSTM $\eta$ = 0.001) | 0.511 (LSTM $\eta$ = 0.001) | 1.458 (LSTM $\eta$ = 0.001) | 3.163 (LSTM $\eta$ = 0.001) |
| | Summer | 0.916 (LSTM-PC $\eta$ = 0.007) | 0.243 (LSTM-PC $\eta$ = 0.007) | 0.722 (LSTM-PC $\eta$ = 0.007) | 1.539 (LSTM-PC $\eta$ = 0.007) |
| | Autumn | 1.98 (ANN $\eta$ = 0.1) | 0.492 (LSTM-PC $\eta$ = 0.005) | 1.413 (LSTM-PC $\eta$ = 0.005) | 3.416 (ANN $\eta$ = 0.1) |
| | Winter | 2.31 (ANN $\eta$ = 0.005) | 0.611 (ANN $\eta$ = 0.5) | 1.695 (ANN $\eta$ = 0.5) | 4.166 (ANN $\eta$ = 0.5) |
| Tongyeong | Spring | 1.583 (ANN $\eta$ = 0.1) | 0.439 (LSTM $\eta$ = 0.005) | 1.26 (LSTM $\eta$ = 0.005) | 2.758 (ANN $\eta$ = 0.1) |
| | Summer | 0.878 (LSTM-PC $\eta = 0.003$) | 0.226 (LSTM-PC $\eta = 0.003$) | 0.671 (LSTM-PC $\eta = 0.003$) | 1.478 (LSTM-PC $\eta = 0.003$) |
| | Autumn | 1.783 (LSTM-PC $\eta = 0.005$) | 0.463 (LSTM-PC $\eta = 0.005$) | 1.336 (LSTM-PC $\eta = 0.005$) | 3.063 (LSTM-PC $\eta = 0.005$) |
| | Winter | 2.144 (ANN $\eta = 0.3$) | 0.585 (ANN $\eta = 0.3$) | 1.635 (ANN $\eta = 0.3$) | 3.840 (ANN $\eta = 0.3$) |
| Jeonju | Spring | 2.281 (LSTM-PC $\eta = 0.005$) | 0.622 (DNN $\eta = 0.1$) | 1.779 (DNN $\eta = 0.1$) | 3.983 (LSTM-PC $\eta = 0.005$) |
| | Summer | 1.246 (LSTM-PC $\eta = 0.001$) | 0.324 (LSTM-PC $\eta = 0.003$) | 0.961 (LSTM-PC $\eta = 0.003$) | 2.090 (LSTM-PC $\eta = 0.001$) |
| | Autumn | 2.22 (LSTM $\eta = 0.009$) | 0.584 (LSTM-PC $\eta = 0.007$) | 1.675 (LSTM-PC $\eta = 0.007$) | 3.837 (LSTM $\eta = 0.009$) |
| | Winter | 2.601 (ANN $\eta = 0.3$) | 0.716 (ANN $\eta = 0.3$) | 1.984 (ANN $\eta = 0.3$) | 4.699 (ANN $\eta = 0.3$) |

**Table 3.**

Values of RMSE, MAPE, MAE and Theil's-U of ten metropolitan cities in four seasons in testing 2, where $\eta$ denotes the learning rate.

| City | Season | RMSE | MAPE | MAE | Theil's-U |
|---|---|---|---|---|---|
| Seoul | Spring | 13.302 (LSTM-PC $\eta = 0.001$) | 22.944 (LSTM-PC $\eta = 0.009$) | 10.7 (LSTM-PC $\eta = 0.009$) | 0.126 (LSTM-PC $\eta = 0.001$) |
| | Summer | 9.399 (LSTM-PC $\eta = 0.003$) | 11.087 (LSTM-PC $\eta = 0.007$) | 7.226 (LSTM $\eta = 0.009$) | 0.071 (LSTM-PC $\eta = 0.003$) |
| | Autumn | 11.313 (LSTM-PC $\eta = 0.001$) | 14 (LSTM $\eta = 0.007$) | 8.055 (LSTM $\eta = 0.007$) | 0.091 (LSTM-PC $\eta = 0.001$) |
| | Winter | 11.054 (LSTM-PC $\eta = 0.001$) | 14.915 (LSTM-PC $\eta = 0.001$) | 8.531 (LSTM-PC $\eta = 0.001$) | 0.095 (LSTM-PC $\eta = 0.009$) |
| Daejeon | Spring | 11.468 (LSTM-PC $\eta = 0.001$) | 17.095 (LSTM-PC $\eta = 0.001$) | 9.64 (LSTM-PC $\eta = 0.001$) | 0.094 (LSTM-PC $\eta = 0.001$) |
| | Summer | 7.46 (ANN $\eta = 0.1$) | 7.433 (LSTM $\eta = 0.009$) | 5.737 (ann $\eta = 0.1$) | 0.048 (ANN $\eta = 0.1$) |
| | Autumn | 7.827 (LSTM-PC $\eta = 0.001$) | 7.99 (LSTM-PC $\eta = 0.005$) | 5.809 (LSTM-PC $\eta = 0.001$) | 0.051 (LSTM-PC $\eta = 0.001$) |
| | Winter | 10.403 | 11.352 | 7.862 | 0.074 |
| | | (LSTM $\eta = 0.001$) | (LSTM-PC $\eta = 0.001$) | (LSTM-PC $\eta = 0.001$) | (LSTM-PC $\eta = 0.007$) |
| Daegu | Spring | 12.798 (LSTM $\eta = 0.001$) | 21.951 (LSTM-PC $\eta = 0.003$) | 10.176 (LSTM-PC $\eta = 0.003$) | 0.121 (LSTM $\eta = 0.001$) |
| | Summer | 9.307 (ANN $\eta = 0.1$) | 9.392 (LSTM-PC $\eta = 0.007$) | 6.996 (LSTM-PC $\eta = 0.007$) | 0.064 (ANN $\eta = 0.1$) |
| | Autumn | 9.614 (LSTM-PC $\eta = 0.003$) | 10.853 (LSTM-PC $\eta = 0.003$) | 7.504 (LSTM $\eta = 0.001$) | 0.067 (LSTM $\eta = 0.001$) |
| | Winter | 13.404 (LSTM $\eta = 0.001$) | 16.484 (LSTM-PC $\eta = 0.001$) | 9.594 (LSTM-PC $\eta = 0.001$) | 0.114 (LSTM-PC $\eta = 0.005$) |
| Busan | Spring | 12.085 (LSTM $\eta = 0.001$) | 19.212 (LSTM $\eta = 0.001$) | 9.893 (LSTM $\eta = 0.001$) | 0.101 (LSTM-PC $\eta = 0.003$) |
| | Summer | 7.166 (ANN $\eta = 0.1$) | 7.147 (ANN $\eta = 0.1$) | 5.708 (ann $\eta = 0.1$) | 0.045 (ANN $\eta = 0.1$) |
| | Autumn | 10.032 (LSTM-PC $\eta = 0.009$) | 12.081 (LSTM $\eta = 0.007$) | 7.744 (LSTM $\eta = 0.007$) | 0.072 (LSTM $\eta = 0.005$) |
| | Winter | 14.173 (LSTM-PC $\eta = 0.001$) | 18.804 (LSTM-PC $\eta = 0.003$) | 10.081 (LSTM-PC $\eta = 0.001$) | 0.131 (ANN $\eta = 0.007$) |
| Incheon | Spring | 12.629 (ANN $\eta = 0.5$) | 17.822 (ANN $\eta = 0.3$) | 9.989 (ANN $\eta = 0.3$) | 0.097 (ANN $\eta = 0.5$) |
| | Summer | 8.864 (LSTM-PC $\eta = 0.005$) | 9.755 (LSTM-PC $\eta = 0.005$) | 6.969 (LSTM-PC $\eta = 0.005$) | 0.057 (LSTM-PC $\eta = 0.005$) |
| | Autumn | 10.941 (LSTM-PC $\eta = 0.005$) | 14.308 (LSTM-PC $\eta = 0.005$) | 8.64 (LSTM-PC $\eta = 0.005$) | 0.083 (LSTM-PC $\eta = 0.005$) |
| | Winter | 9.832 (LSTM-PC $\eta = 0.009$) | 11.892 (LSTM-PC $\eta = 0.003$) | 7.585 (LSTM-PC $\eta = 0.003$) | 0.078 (LSTM-PC $\eta = 0.009$) |
| Gwangju | Spring | 13.862 (LSTM $\eta = 0.001$) | 19.23 (LSTM $\eta = 0.001$) | 11.248 (LSTM $\eta = 0.001$) | 0.109 (LSTM $\eta = 0.001$) |
| | Summer | 7.093 (ANN $\eta = 0.7$) | 6.659 (LSTM-PC $\eta = 0.007$) | 5.53 (LSTM-PC $\eta = 0.007$) | 0.044 (ANN $\eta = 0.7$) |
| | Autumn | 8.895 (LSTM $\eta = 0.003$) | 10.391 (LSTM $\eta = 0.003$) | 6.945 (LSTM $\eta = 0.003$) | 0.059 (LSTM $\eta = 0.003$) |
| | Winter | 12.703 (LSTM-PC $\eta = 0.005$) | 15.463 (ANN $\eta = 0.5$) | 10.008 (LSTM-PC $\eta = 0.005$) | 0.096 (LSTM-PC $\eta = 0.009$) |
| Pohang | Spring | 14.322 (LSTM $\eta = 0.001$) | 24.717 (LSTM-PC $\eta = 0.003$) | 11.793 (LSTM $\eta = 0.001$) | 0.123 (LSTM $\eta = 0.001$) |
| | Summer | 7.99 (ANN $\eta = 0.7$) | 8.122 (LSTM $\eta = 0.001$) | 6.326 (LSTM $\eta = 0.001$) | 0.051 (ANN $\eta = 0.9$) |
| | Autumn | 9.408 (ANN $\eta = 0.3$) | 11.043 (ANN $\eta = 0.3$) | 7.598 (ANN $\eta = 0.3$) | 0.065 (ANN $\eta = 0.3$) |
| | Winter | 12.832 (ANN $\eta = 0.5$) | 17.431 (LSTM $\eta = 0.007$) | 9.866 (ANN $\eta = 0.5$) | 0.109 (ANN $\eta = 0.5$) |
| Mokpo | Spring | 11.435 (ANN $\eta = 0.7$) | 15.024 (ANN $\eta = 0.7$) | 9.502 (LSTM-PC $\eta = 0.007$) | 0.082 (ANN $\eta = 0.7$) |
| | Summer | 5.839 (ANN $\eta = 0.1$) | 5.525 (LSTM $\eta = 0.007$) | 4.451 (LSTM $\eta = 0.007$) | 0.036 (ANN $\eta = 0.1$) |
| | Autumn | 6.899 (LSTM $\eta = 0.005$) | 7.702 (ANN $\eta = 0.3$) | 5.494 (ANN $\eta = 0.1$) | 0.046 (LSTM $\eta = 0.003$) |
| | Winter | 8.16 (ANN $\eta = 0.1$) | 9.248 (ANN $\eta = 0.1$) | 6.54 (ANN $\eta = 0.1$) | 0.058 (ANN $\eta = 0.1$) |
| Tongyeong | Spring | 10.479 (LSTM $\eta = 0.001$) | 13.926 (LSTM-PC $\eta = 0.003$) | 8.613 (LSTM-PC $\eta = 0.003$) | 0.076 (LSTM $\eta = 0.001$) |
| | Summer | 6.549 (LSTM $\eta = 0.009$) | 6.166 (LSTM $\eta = 0.009$) | 5.002 (LSTM $\eta = 0.009$) | 0.039 (LSTM $\eta = 0.005$) |
| | Autumn | 8.63 (ANN $\eta = 0.1$) | 9.729 (LSTM $\eta = 0.007$) | 6.747 (LSTM-PC $\eta = 0.003$) | 0.059 (ANN $\eta = 0.1$) |
| | Winter | 12.371 (LSTM $\eta = 0.001$) | 14.826 (LSTM-PC $\eta = 0.003$) | 9.15 (LSTM-PC $\eta = 0.001$) | 0.101 (ANN $\eta = 0.7$) |
| Jeonju | Spring | 12.011 (LSTM-PC $\eta = 0.003$) | 14.698 (LSTM $\eta = 0.001$) | 9.658 (LSTM $\eta = 0.001$) | 0.088 (LSTM-PC $\eta = 0.003$) |
| | Summer | 7.027 (ANN $\eta = 0.9$) | 6.351 (LSTM $\eta = 0.007$) | 5.412 (LSTM-PC $\eta = 0.005$) | 0.041 (ANN $\eta = 0.9$) |
| | Autumn | 7.386 (LSTM-PC $\eta = 0.001$) | 8.178 (LSTM-PC $\eta = 0.001$) | 5.98 (LSTM-PC $\eta = 0.001$) | 0.047 (LSTM-PC $\eta = 0.1$) |
| | Winter | 8.899 (ANN $\eta = 0.1$) | 11.033 (ANN $\eta = 0.1$) | 6.992 (ANN $\eta = 0.1$) | 0.066 (ANN $\eta = 0.1$) |

Figs. 6-9 depict the lowest RMSE, MAPE, MAP, and Theil's-U of ten metropolitan cities for all three training (2500, 5000, 7500) epochs in spring, summer, autumn, winter in testing 1 and 2, respectively. Tables 2 and 3 present the comparison of the RMSE, MAPE, MAE and Theil's-U statistics in four seasons of ten metropolitan cities in testing 1 and 2, respectively.

From testing 1, as given in Table 2, the RMSE (MAPE) of ANN (LSTM) has a lowest value of 1.583 (0.429) at $\eta = 0.1$ (0.005) in spring in Tongyeong. The RMSE (MAE) of LSTM-PC has a lowest value of 0.878 (0.243) at $\eta = 0.003$ (0.007) in summer in Tongyeong. In autumn of Busan, the RMSE (MAE) value of LSTM-PC is a lowest value 1.72 (1.298) at $\eta = 0.005$. In winter of Daegue, the MAPE (Theil's-U) of ANN has a value of 1.509 ($3.694 \times 10^{-3}$) at $\eta = 0.3$ lowest than that of other metropolitan cities.

As given in Table 3, from testing 2, in spring of Tongyeong, the RMSE (MAPE) of LSTM (LSTM-PC) has a lowest value of 10.479 (13.926) at $\eta = 0.001$ (0.003). The RMSE (MAE) of LSTM-PC (ANN) has a lowest value of 5.839 (4.451) at learning rate $\eta = 0.005$ (0.1) in Summer in Mokpo. In autumn of Mokpo, the RMSE (MAE) value of LSTM (ANN) is a lowest value 6.899 (5.494) at $\eta = 0.005$ (0.1). It is not good accuracy characteristically in autumn of testing 2, but in autumn, the RMSE value of DNN is 8.076 at $\eta = 0.1$ in Jeonju, while that of ELM is 8.196 in Mokpo for 5000 training epochs. In winter of Mokpo, the MAPE (Theil's-U) of ANN has a value of 6.54 (0.057) at $\eta = 0.1$ lowest than that of other metropolitan cities.

In the architecture of our studies, we allow our way to simulate the DNN, which makes optimal values for selecting relatively the number of input and hidden units. It is considered that the LSTM case tries the computer-simulation having a pertinent number of input and recurrent units.

## 4. Summary

This paper is dedicated to the problems of predicting the daily (low frequency) average temperature and humidity of ten metropolitan cities (Seoul, Daejeon, Daegu, Busan, Incheon, Gwangju, Pohang, Mokpo, Tongyeong, and Jeonju) in Korea, using five neural network machine learning methods. We extracted the data of the manned regional meteorological offices of the KMA to ensure the reliability of data, and these are for seven years from 2014 to 2020. We set the five learning rate values for the ANN and the DNN as 0.1, 0.3, 0.5, 0.7, and 0.9, while those for LSTM and LSTM-PC are set to 0.001, 0.003, 0.005, 0.007, 0.009 for 2500, 5000, 7500 training epochs. The predicted values of the ELM are also obtained by averaging the results trained 2500, 5000, and 7500 epochs. From the result of outputs, the RMSE, MAPE, MAE, and Theil's-U statistics are simulated for performance evaluation, and we have compared these statistics after manipulating of five neural network methods.

We have found our results as follows: In testing 1 (the temperature predicted in the input layer with four input nodes), the RMSE value is 1.583 for the 7500 training epochs of the ANN at learning rate $\eta_1 = 0.1$ in spring in Tongyeong. The RMSE value of the LSTM-PC at learning rate $\eta = 0.003$ in summer in Tongyeong is 0.878 for 2500 training epochs, while that for the LSTM-PC at $\eta = 0.005$ is 1.72 for 5000 training epochs in autumn in Busan. The RMSE value of the LSTM-PC ($\eta = 0.007$) is 2.078 for 5000 training epochs in winter in Daegu. Among the four seasons, the LSTM-PC shows good performance in three seasons (summer, autumn, and winter). Particularly, when the LSTM ($\eta = 0.003$) is trained 2500 training epochs in summer in Tongyeong, the RMSE has the smallest value with 0.878. In testing 2 (the humidity predicted in the input layer with four input nodes), In spring, the RMSE value of the LSTM ($\eta = 0.001$) is 10.609 for the 7500 training epochs in spring in Tongyeong. The RMSE value of the ANN ($\eta = 0.1$) is 5.839 for 2500 training epochs in summer in Mokpo. The RMSE of the LSTM ($\eta = 0.005$) was 6.891 for 7500 traning epochs in autumn in Mokpo, The RMSE value is 8.16 for 2500 training epochs of the ANN ($\eta = 0.1$) in winter in Mokpo. When the ANN ($\eta = 0.1$) is trained 2500 times in the summer in Mokpo, the RMSE has the smallest value with 5.839.

In fact, to predict the temperature, the LSTM-PC in testing 1 has the lowest value in the autumn in Busan for 5000 training epoch. In winter, the LSTM-PC shows the smallest error for 5000 training epochs in winter in Daegu. The RMSE of the LSTM in testing 2 has the lowest value for 7500 training epochs in autumn in Mokpo. For the humidity prediction in the summer in Mokpo, the RMSE of LSTM is shown the lowest value of 5.732. Finally, in the temperature and humidity predictions, the RMSEs are the lowest in summer in Mokpo.

Noteworthy, our result provides the evidence that the LSTM is an effective method of predicting one meteorological factor (temperature) rather than the DNN. Prediction accuracy of temperature among our result is approximately consistent to the result obtained from Chinese stock data via the deep learning network method [34]. A detailed simulation and analysis for neural network method with the different number of input nodes will be presented elsewhere [93].

Presently, researchers have not exerted more investigations and analysis with high and low frequency time-series data in meteorological and climatological fields, but we are due to study and explore continuously meteorological factors with high frequency time-series data in the future if accumulating more high frequency time-series data. With the detailed understanding of the interplay between temperature and humidity [94], we hope that the researches may be extended and improved to treat the neural network methods, if there are the robust correlations between other types of meteorological and climatological factors. We implicitly consider that there is a need to expand and study with extant big data for the past five decades (and over) than the recent data. We have presently limited ourselves to the relevant neural network methods in only ten metropolitan cities, but our study can provide a valuable insight on how more machine learning methods are expanded and analyzed to other cities of Korea and metropolitan cities of world.

## Acknowledgments

This is supported by a Research Grant of Pukyong National University (2021 year).